\documentclass[10pt]{article}
\usepackage{graphicx}
\usepackage{cite}
\usepackage{amssymb,amsfonts,amsmath}

\newcommand{\be}{\begin{eqnarray}}
\newcommand{\ee}{\end{eqnarray}}
\newcommand{\rmc}{{\rm C}}
\newcommand{\rmd}{{\rm D}}
\usepackage[labelfont=bf,labelsep=period,justification=raggedright]{caption}

\begin{document}

\title{Evolutionary instability of Zero Determinant strategies demonstrates that winning isn't everything}

\author{ Christoph Adami$^{1,2,3,\star}$ and Arend Hintze$^{1,3}$ \\ \\
$^1$Department of Microbiology and Molecular Genetics\\
$^2$Department of Physics and Astronomy\\
$^3$BEACON Center for the Study of Evolution in Action\\
Michigan State University, East Lansing, MI 48824\\
$\ast$ E-mail: adami@msu.edu}

\maketitle


\begin{abstract} Zero Determinant  strategies are a new class of probabilistic and conditional strategies that are able to unilaterally set the expected payoff of an opponent in iterated plays of the Prisoner's Dilemma irrespective of the opponent's strategy (coercive strategies), or else to set the ratio between the player's and their opponent's expected payoff (extortionate strategies). Here we show that Zero Determinant strategies are at most weakly dominant, are not evolutionarily stable, and will instead evolve into less coercive strategies. We show that Zero Determinant strategies with an informational advantage over other players that allows them to recognize each other can be evolutionarily stable (and able to exploit other players). However, such an advantage is bound to be short-lived as opposing strategies evolve to counteract the recognition. \end{abstract}

\vskip 1cm
Evolutionary Game Theory (EGT) has been around for over 30 years, but apparently the theory still has surprises up its sleeve. Recently, Press and Dyson discovered a new class of strategies within the realm of two-player iterated games that allow one player to unilaterally set the opponent's payoff, or else extort the opponent to accept an unequal share of payoffs~\cite{PressDyson2012,StewartPlotkin2012}. This new class of strategies, named ``Zero Determinant" (ZD) strategies, exploits a curious mathematical property of the expected payoff for a stochastic conditional ``memory-one" strategy.  In the standard game of EGT called the Prisoner's Dilemma (PD), the possible moves are termed ``Cooperate" (C) and ``Defect" (D) (as the original objective of evolutionary game theory was to understand the evolution of cooperation~\cite{AxelrodHamilton1981, MaynardSmith1982,HofbauerSigmund1998}), and the payoffs for this game can be written in terms of a payoff matrix
\be
\bordermatrix{~ & \rmc & \rmd \cr
                  \rmc & R & S \cr
                  \rmd & T & P \cr}\;,\label{pay}
 \ee
where the payoff is understood to be given to the ``row" player (e.g., for the pair of plays ``DC", the player who chose D obtains payoff $T$ while the opponent receives payoff $S$).

As opposed to deterministic strategies such as ``Always Cooperate" or ``Always Defect", stochastic strategies are defined by probabilities to engage in one move or the other. ``Memory-one" strategies make their move depending on theirs as well as their opponent's last move: perhaps the most famous of all memory-one strategies within the iterated Prisoner's Dilemma (IPD) game called ``Tit-for Tat" plans its moves as a function of only its opponent's last move. Memory-one probabilistic strategies are defined by four probabilities, namely to cooperate given the four possible outcomes of the last play. While probabilistic one-memory iterated games were studied as early as 1990~\cite{Nowak1990,NowakSigmund1990} and more recently by us~\cite{Iliopoulosetal2010}, the existence of ZD strategies still took the field by surprise (even though such strategies had in fact been discovered earlier~\cite{Boerlijstetal1997,Sigmund2010}) . 

Stochastic conditional strategies are defined by a set of four probabilities $p_1,p_2,p_3,p_4$ to cooperate given that the last encounter between this player and his opponent resulted in the outcome CC ($p_1$),CD ($p_2$), DC ($p_3$) or DD ($p_4$). ZD strategies are defined by fixing $p_2$ and $p_3$ to be a very specific function of $p_1$ and $p_4$ as described in Methods. Press and Dyson show that when playing against the ZD strategy, the payoff that an opponent O reaps is defined by the payoffs (\ref{pay}) and only the two remaining probabilities $p_1$ and $p_4$ ($p_2$ and $p_3$ being fixed by Eq.~(\ref{zdprobs})) 
\be
E({\rm O,ZD})=f(\vec p)=\frac{(1-p_1)P+p_4R}{1-p_1+p_4}\;, \label{zd1}
\ee
while the the ZD strategist's payoff against O 
\be
E({\rm ZD,O)}=g(\vec p,\vec q) \label{zd2}
\ee
is still a complicated function of both ZD's and O's strategy that is too lengthy to write down here [but see Methods for the mean payoff given the standard~\cite{AxelrodHamilton1981} PD values  $(R,S,T,P)=(3,0,5,1)$]. 

In Eqs.~(\ref{zd1},\ref{zd2}) we adopted the notation of a payoff matrix where the payoff is given to the ``row-player". Note that the payoff that the ZD player forces upon its opponent is not necessarily smaller than what the ZD player receives. For example,
the payoff for ZD against the strategy ``All-D" that defects unconditionally at every move $\vec q=(0,0,0,0)$ is
\be E(\textrm{ZD, All-D})=P+\frac {p_4(R-P)}{1-p_1+p_4}\left(\frac{R-S}{P-T}\right)\;, \label{alld}
\ee
which is strictly lower than (\ref{zd1}) for all games in the realm of the PD parameters.

Interestingly, a ZD strategist can also extort an unfair share of the payoffs from the opponent, who however could refuse it (turning the game into a version of the Ultimatum Game~\cite{Nowaketal2000}). In extortionate games, the strategy being preyed upon can increase their own payoff by modifying their own strategy $\vec q$, but this only increases the extortionate strategy's payoff. As a consequence, Press and Dyson conclude that a ZD strategy will always dominate any opponent that adapts its own strategy to maximize their payoff, for example by Darwinian evolution~\cite{PressDyson2012}. 

Here we show that ZD strategies (those who fix the opponent's payoff as well as those who extort) are actually evolutionarily unstable, are easily outcompeted by fairly common strategies, and quickly evolve to become non-ZD strategies. However, if ZD strategies can determine who they are playing against (either by recognizing a tag or by analyzing the opponent's response), ZD strategists are likely to be very powerful agents against unwitting opponents.

\section*{Results}
\subsubsection*{Effective payoffs and Evolutionarily Stable Strategies}
In order to determine whether a strategy will succeed in a population, Maynard Smith proposed the concept of an ``Evolutionarily Stable Strategy" (or ESS)~\cite{MaynardSmith1982}. For a game involving arbitrary strategies $I$ and $J$, the ESS is easily determined by an inspection of the payoff matrices of the game as follows: $I$ is an ESS if the payoff $E(I,I)$ when playing itself is larger than the payoff $E(J,I)$ between any other strategy $J$ and $I$, i.e., 
$I\ {\rm is}\  {\rm ESS}\  {\rm if}\  E(I,I)>E(J,I)$. In case $ E(I,I)=E(J,I)$, then $I$ is an ESS if at the same time
 $E(I,J)>E(J,J)$. These equations teach us a fundamental lesson in evolutionary biology: it is not sufficient for a strategy to outcompete another strategy in direct competition, that is, winning isn't everything. Rather, a strategy must also play well against itself. The reason for this is that if a strategy plays well against an opponent but reaps less of a benefit competing against itself, then it will be able to invade a population but will quickly have to compete against its own offspring and its rate of expansion slows down. This is even more pronounced in populations with a spatial structure, where offspring are placed predominantly close to the progenitor. If the competing strategy in comparison plays very well against itself, then a strategy that only plays well against an opponent may not even be able to invade. 

If we assume that two opponents play a sufficiently large number of games, their average payoff approaches the payoff of the Markov stationary state~\cite{HauertSchuster1997,PressDyson2012}. We can use this mean expected payoff as the payoff to be used in the payoff matrix $E$ that will determine the ESS. For ZD strategies playing O (other) strategies, we know that ZD enforces $E$(O,ZD)=$f(\vec p)$ shown in equation (\ref{zd1}), while ZD receives $g(\vec p,\vec q)$ [Eq.~(\ref{zd2})]. But what are the diagonal entries in this matrix? We know that ZD enforces the score (\ref{zd1})  {\em regardless} of the opponent's strategy, which implies that it also enforces this on another ZD strategist. Thus, $E$(ZD,ZD)=$f(\vec p)$. The payoff of O against itself only depends on O's strategy $\vec q$: $E$(O,O)=$h(\vec q)$, and is the key variable in the game once the ZD strategy is fixed. The effective payoff matrix then becomes
\be 
E=\begin{pmatrix} 
      f(\vec p) & g(\vec p,\vec q)\\
      f(\vec p)& h(\vec q)\\
   \end{pmatrix}\;.  \label{zdmat} \ee
Note that by writing mean expected payoffs into a matrix such as (\ref{zdmat}), we have effectively defined a new game in which the possible {\em moves} are ZD and O. It is in terms of these moves that we can now consider dominance and evolutionary stability of these strategies.  

The payoff matrix of any game can be brought into a normalized form with vanishing diagonals without affecting the competitive dynamics of the strategies by subtracting a constant term from each column~\cite{Zeeman1980}, so the effective payoff (\ref{zdmat}) is equivalent to
\be 
E=\begin{pmatrix} 
   0 & g(\vec p,\vec q)-h(\vec q)\\
    0& 0\\
   \end{pmatrix}\;.  \label{zdmat1} \ee
We notice that the fixed payoff $f(\vec p)$ has disappeared, and the winner of the competition is determined entirely from the sign of $g(\vec p,\vec q)-h(\vec q)$, as seen by an inspection of the ESS equations. If $g(\vec p,\vec q)-h(\vec q)>0$, the ZD strategy is a weak ESS (see, e.g.,~\cite{Nowak2006}, p. 55). If $g(\vec p,\vec q)-h(\vec q)<0$, the opposing strategy is the ESS. In principle, a mixed strategy (a population mixture of two strategies that are in equilibrium) can be an ESS~\cite{MaynardSmith1982} but this is not possible here precisely because ZD enforces the same score on others as it does on itself. 

\subsubsection*{Evolutionary dynamics of ZD via replicator equations} 

\begin{figure}[!t] 
   \centering
   \includegraphics[width=4in]{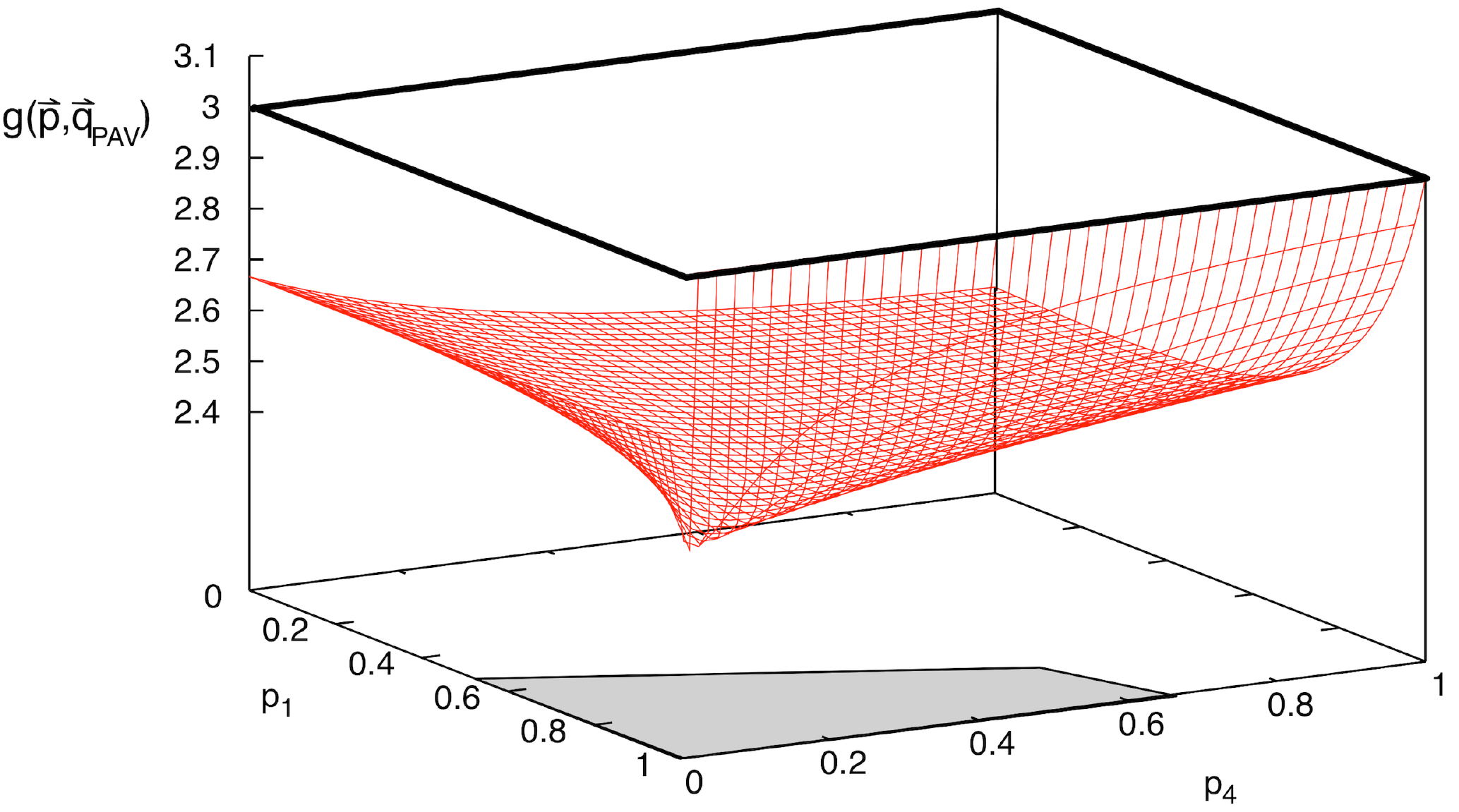} 
   \caption{{\bf Mean expected payoff of arbitrary ZD strategies playing ``Pavlov".} The payoff $g(\vec p, \vec q_{\rm PAV})$  (red surface) defined by the allowed set $(p_1,p_4)$ (shaded region) against the strategy Pavlov, given by the probabilities $\vec q_{\rm PAV}=(1,0,0,1)$. Because $g(\vec p, \vec q_{\rm PAV})$ is everywhere smaller than $h(\vec q_{\rm PAV})=3$ (except on the line $p_1=1$), it is Pavlov which is the ESS for all allowed values   $(p_1,p_4)$, according to Eq.~(\ref{zdmat1}). For $p_1=1$, ZD and Pavlov are equivalent as the entire payoff matrix~(\ref{zdmat1}) vanishes (even though the strategies are not the same).}
   \label{fig:example}
\end{figure}

ZD strategies are defined by setting two of the four probabilities in a strategy to specific values so that the payoff $E$(O,ZD) depends only on the other two, but not on the strategy O. In Ref.~\cite{PressDyson2012}, the authors chose to fix $p_2$ and $p_3$, leaving $p_1$ and $p_4$ to define a family of ZD strategies. The requirement of a vanishing determinant limits the possible values of $p_1$ (the probability to cooperate if in the previous move both players cooperated) to close to 1, while $p_4$ must be near (but not equal to) zero.
Let us study an example ZD strategy defined by the values $p_1=0.99$ and $p_4=0.01$. The results we present do not depend on the choice of the ZD strategy. An inspection of Eq.~(\ref{zd1}) shows that, if we use the standard payoffs of the Prisoner's Dilemma $(R,S,T,P)=(3,0,5,1)$, then $f(\vec p)=2$. If we study the strategy ``All-D" (always defect, defined by the strategy vector $\vec q=(0,0,0,0)$ as opponent) we find that $g(\vec p,\vec q)=0.75$, while $h(\vec q)=1$. As a consequence, $g(\vec p,\vec q)-h(\vec q)$ is negative and All-D is the ESS, that is, ZD will lose the evolutionary competition with All-D. However, this is not surprising as ZD's payoff against All-D is in fact lower than what ZD forces All-D to accept, as mentioned earlier. But let us consider the competition of ZD with the strategy ``Pavlov", which is a strategy of ``win-stay-lose-shift" that can outperform the well-known ``Tit-for-Tat" strategy in direct competition~\cite{NowakSigmund1993}. Pavlov is given by the strategy vector $\vec q_{\rm PAV}=(1,0,0,1)$, which (given the ZD strategy $\vec p$ described above and the standard payoffs) returns E(ZD,PAV)=11/27$\approx 2.455$ to the ZD player, while Pavlov is forced to receive $f(\vec p)=2$. Thus, ZD {\em wins every direct competition} with Pavlov. Yet, Pavlov is the ESS because it cooperates with itself, so $h(\vec q)=3$. We show in Figure~\ref{fig:example} that the dominance of Pavlov is not restricted to the example ZD strategy $p_1=0.99$ and $p_4=0.01$, but holds for all possible ZD strategies within the allowed space.

\begin{figure}[htbp] 
    \begin{center}
 \includegraphics[width=3.0in]{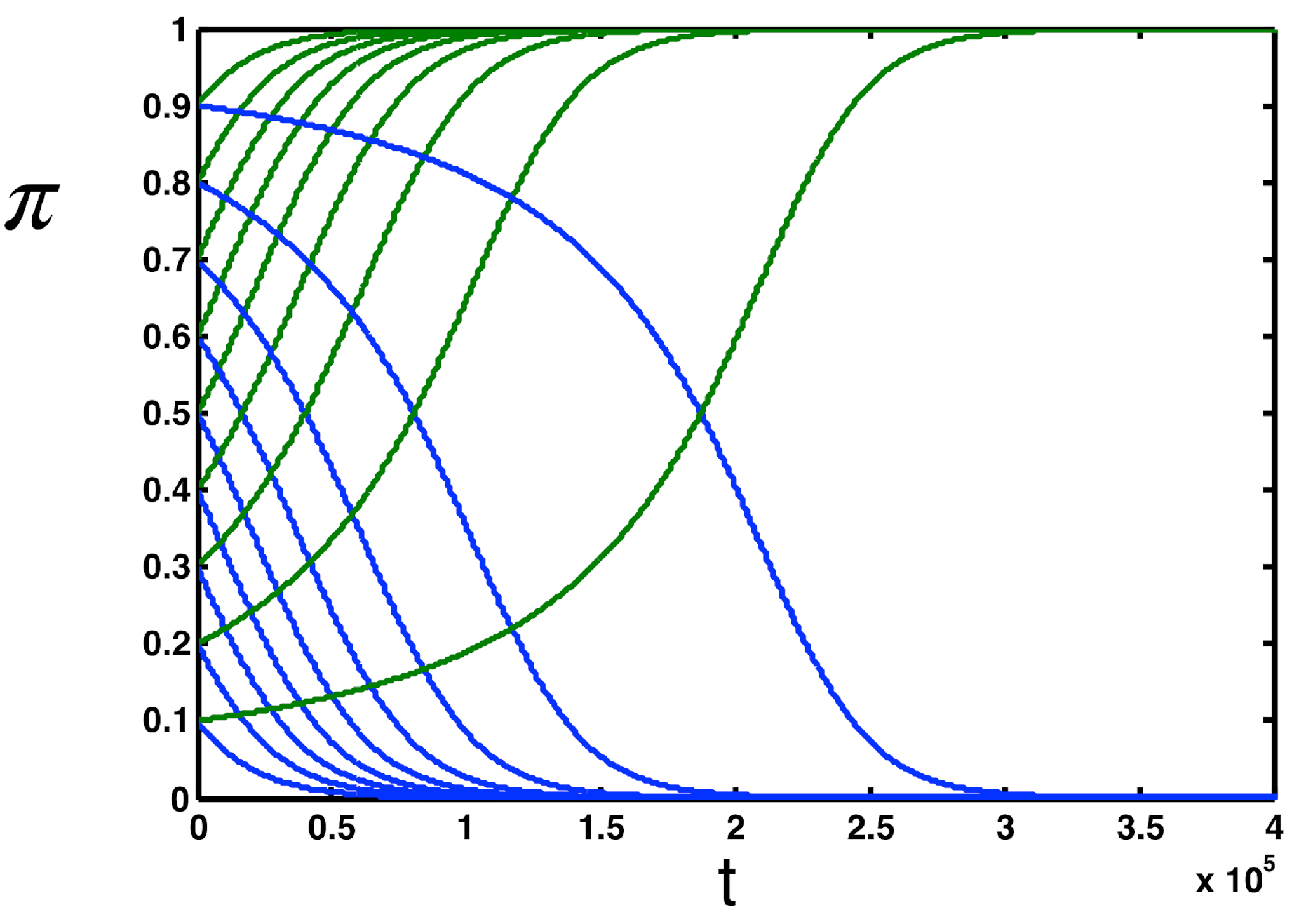} 
    \caption{{\bf Population fractions of ZD vs. Pavlov over time.} Population fractions $\pi_{\rm ZD}$ (blue) and $\pi_{\rm PAV}$ (green) as a function of time for initial ZD concentrations $\pi_{\rm ZD}(0)$ between 0.1 and 0.9. } 
    \label{fig:alld}
    \end{center}
 \end{figure}

We can check that Pavlov is the ESS by following the population fractions as determined by the replicator equations~\cite{TaylorJonker1978,Zeeman1980,HofbauerSigmund1998}, which describe the frequencies of strategies in a population  
\be
\dot \pi_i=\pi_i(w_i-\bar w), \label{rk}
\ee
 where $\pi_i$ is the population fraction of strategy $i$, $w_i$ is the fitness of strategy $i$, and $\bar w$ is the average fitness in the population. In our case, the fitness of strategy $i$ is the mean payoff for this strategy, so
 \begin{eqnarray}
 w_{\rm ZD}&=&\pi_{\rm ZD}E({\rm ZD,ZD})+\pi_{\rm O} E({\rm ZD,O})\\
 w_{\rm O}&=&\pi_{\rm ZD}E({\rm O,ZD})+\pi_{\rm O} E({\rm O,O})
 \end{eqnarray}
 and $\bar w=\pi_{\rm ZD} w_{\rm ZD}+\pi_{\rm O}w_{\rm O}$. We show $\pi_{\rm ZD}$ and $\pi_{\rm AllD}$ (with $\pi_{\rm ZD}+\pi_{\rm AllD}=1$) in Fig.~\ref{fig:alld} as a function of time for different initial conditions, and confirm that Pavlov drives ZD to extinction regardless of initial density. 
 
\subsubsection*{Evolutionary dynamics of ZD in agent-based simulations}
 
 It could be argued that an analysis of evolutionary stability within the replicator equations ignores the complex game play that occurs in populations where the payoff is determined in each game, and where two strategies meet by chance and survive based on their accumulated fitness. We can test this by following ZD strategies in contest with Pavlov in an agent-based simulation with a fixed population size of $N_{\rm pop}=1,024$ agents, a fixed replacement rate of 0.1\%  and using a fitness-proportional selection scheme (a death-birth Moran process, see Methods). 
 \begin{figure}[htbp] 
    \centering
    \includegraphics[width=3in]{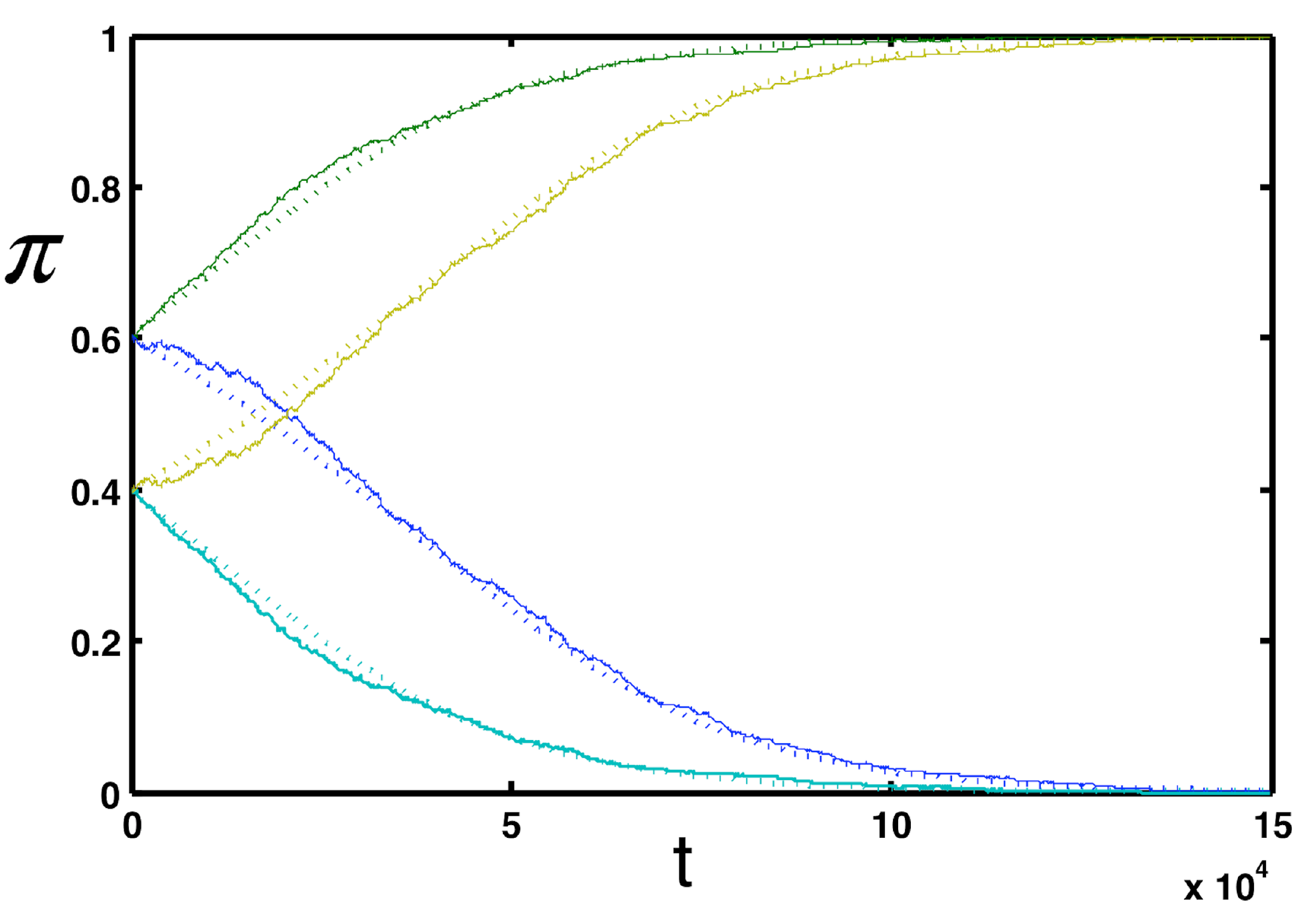} 
    \caption{{\bf Population fractions using agent-based simulations and replicator equations.}
    Population fractions $\pi_{\rm ZD}$ (blue tones) and $\pi_{\rm PAV}$ (green tones) for two different initial concentrations. The solid lines show the average of the population fraction from 40 agent-based simulations as a function of evolutionary time measured in updates, while the dashed lines show the corresponding replicator equations. Because time is measured differently in agent-based simulations as opposed to the replicator equations, we applied an overall scale to the time variable of the Runge-Kutta simulation of Eq.~(\ref{rk}) to match the agent-based simulation.
    \label{fig:sdvszd}}
  \end{figure}

In Fig.~\ref{fig:sdvszd} we show the population fractions $\pi_{\rm ZD}$ and $\pi_{\rm PAV}$ for two different initial conditions [$\pi_{\rm ZD}(0)=0.4$ and 0.6], using a full agent-based simulation (solid lines) or using the replicator equations (dotted lines). While the trajectories differ in detail (likely because in the agent-based simulations generations overlap, the number of encounters is not infinite but dictated by the replacement rate, and payoffs are accumulated over eight opponents randomly chosen from the population),  the dynamics are qualitatively the same. (This can also be shown for any other stochastic strategy $\vec q$ playing against a ZD strategy.) Note that in the agent-based simulations, strategies have to play the first move unconditionally. In the results presented in Fig.~\ref{fig:sdvszd}, we have set this ``first move" probability to $p_C=0.5$ for both Pavlov and ZD. 
 
Agent-based simulations thus corroborate what the replicator equations have already told us, namely that ZD strategies have a hard time surviving in populations because they suffer from the same low payoff that they impose on other strategies if faced with their own kind. However, ZD can win some battles, in particular against strategies that cooperate. For example, the stochastic cooperator GC [``general cooperator", defined by $\vec q = (0.935,0.229,0.266,0.42)$] is the evolutionarily dominating strategy (the fixed point) that evolved at low mutation rates in Ref.~\cite{Iliopoulosetal2010}. GC is a cooperator that is very generous, cooperating after mutual defection almost half the time. GC loses out (in the evolutionary sense) against ZD because $E$(Z,GC)=2.125 while $E$(GC,GC)=2.11 (making ZD a weak ESS), and ZD certainly wins (again in the evolutionary sense) against the unconditional deterministic strategy ``All-C" that always cooperates [see Eq.~(\ref{allc}) in the Methods]. If this is the case, how is it possible that GC is the evolutionary fixed point rather than ZD, when strategies are free to evolve from random ancestors~\cite{Iliopoulosetal2010}?

\subsubsection*{Mutational instability of ZD strategies}

   \begin{figure}[!b] 
   \centering
   \includegraphics[width=3 in]{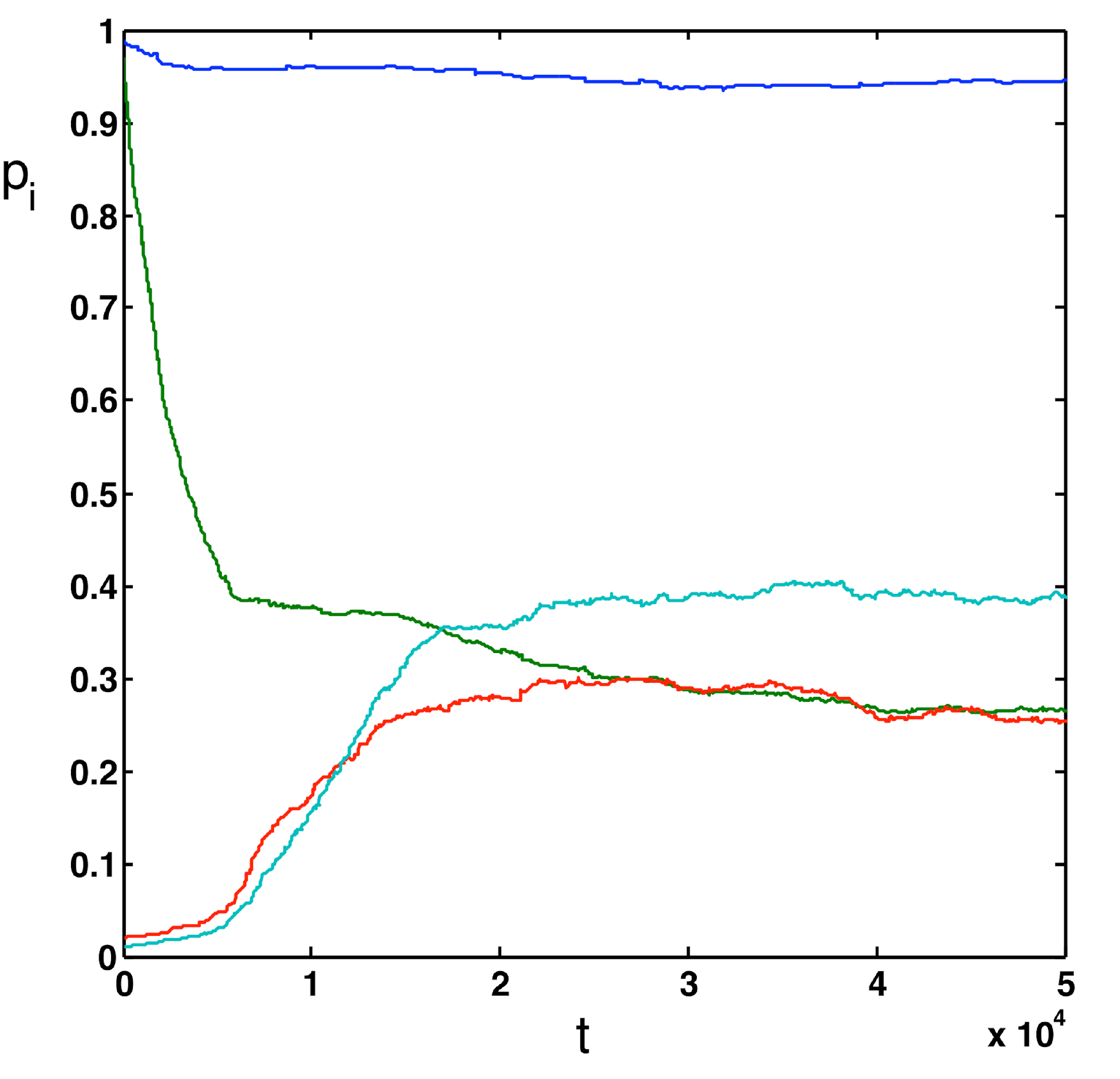} 
   \caption{{\bf Evolution of probabilities on the evolutionary line of descent}. Evolution of probabilities $p_1$ (blue), $p_2$ (green), $p_3$ (red) and $p_4$ (teal) on the evolutionary line of descent of a well-mixed population of 1,024 agents, seeded with the ZD strategy $(p_1,p_2,p_3,p_4)=(0.99, 0.97,0.02,0.01)$. Lines of descent (see Methods) are averaged over 40 independent runs. Mutation rate per gene $\mu=1\%$, replacement rate $r=1\%$. 
   \label{fig:zdevol}}
\end{figure}

To test how ZD fares in a simulation where strategies can evolve (in the previous sections, we only considered the competition between strategies that are fixed), we ran evolutionary (agent-based) simulations in which strategies are encoded genetically. The genome itself evolves via random mutation and fitness-proportional selection. For stochastic strategies, the probabilities are encoded in 5 genes (one unconditional and four conditional probabilities drawn from a uniform distribution when mutated) and evolved as described in the Methods and in Ref.~\cite{Iliopoulosetal2010}. Rather than starting the evolution runs with random strategies, we seeded them with the particular ZD strategy we have discussed here ($p_1=0.99$ and $p_4=0.01$). These simulations show that when we use a mutation rate that favors the strategy GC as the fixed point, ZD evolves into it {\em even though ZD outcompetes GC at zero mutation rate} as we saw in the previous section. In Fig.~\ref{fig:zdevol}, we show the four probabilities that define a strategy over the evolutionary line of descent, followed over 50,000 updates of the population (with a replacement rate of 1\%, this translates on average to 500 generations). The evolutionary line of descent (LOD) is created by taking one of the final genotypes that arose, and following its ancestry backwards mutation by mutation, to arrive at the ZD ancestor used to seed the simulation~\cite{Lenskietal2003}. (Because of the competitive exclusion principle~\cite{Hardin1960}, the individual LODs of all the final genotypes collapse to a single LOD with a fairly recent common ancestor). The LOD confirms what we had found earlier~\cite{Iliopoulosetal2010}, namely that the evolutionary fixed points are independent of the starting strategy and simply reflect the optimal strategy given the amount of uncertainty (here introduced via mutations) in the environment. We thus conclude that ZD is unstable in another sense (besides not being an ESS): it is {\em genetically} or {\em mutationally} unstable, as mutations of ZD are likely {\em not} ZD, and we have shown earlier that ZD generally does not do well against other strategies that defect but are not ZD themselves.

\subsubsection*{Stability of extortionate ZD strategies}

Extortionate ZD strategies (``ZDe" strategies) are those that set the ratio of the ZD strategist's payoff against a non-ZD strategy~\cite{PressDyson2012} rather than setting the opponent's absolute payoff. Against a ZDe strategy, all the opponent can do (in a direct matchup) is to increase their own payoff by optimizing their strategy, but as this increases ZDe's payoff commensurately, the ratio (set by an extortion factor $\chi$, where $\chi=1$ represents a fair game) remains the same. Press and Dyson show that for ZDe strategies with extortion factor $\chi$, the best achievable payoffs for each strategy are [using the conventional IPD values $(R,S,T,P)=(3,0,5,1)$]
\begin{eqnarray}
E({\rm O,ZDe})&=&\frac{12+3\chi}{2+3\chi}\;,\\
E({\rm ZDe,O})&=& \frac{2+13\chi}{2+3\chi}\;,
\end{eqnarray}
which implies that $E({\rm ZDe,O)>E(O,ZDe)}$ for all $\chi>1$. However, ZDe plays terribly against other ZDe strategies, who are defined by a set of probabilities given in Ref.~\cite{PressDyson2012}. Notably, ZDe strategies have $p_4=0$, that is, they never cooperate after both opponents defect. It is easy to show that for $p_4=0$, the mean payoff $E({\rm ZDe,ZDe})=P$ in general, that is, the payoff for mutual defection. As a consequence ZDe can {\em never} be an ESS (not even a weak one) as $E({\rm O,ZDe})>E({\rm ZDe,ZDe})$ for all finite $\chi\geq1$, except when $\chi\to\infty$, where ZDe can be ESS along with an opponent's strategy that has a mean payoff $E({\rm O, O})$ not larger than $P$. We note that the celebrated strategy ``Tit-for-Tat" is technically a ZDe strategy, albeit a trivial one as it always plays fair ($\chi$=1). 

Given that ZD and ZDe are evolutionarily unstable against a large fraction of stochastic strategies, is there no value to this strategy then? We will argue below that strategies that play ZD against non-ZD strategies but a different strategy (for example cooperation) against themselves, may very well be highly fit in the evolutionary sense, and emerge in appropriate evolution experiments.

\subsubsection*{ZD strategies that can recognize other players}

Clearly, winning against your opponents isn't everything if this impairs the payoff against similar or identical strategies. But what if a strategy could recognize who they play against, and switch strategies depending on the nature of the opponent? For example, such a strategy would play ZD against others, but cooperate with other ZD strategists instead. It is in principle possible to design strategies that use a (public or secret) tag to decide between strategies. Riolo et al.~\cite{Rioloetal2001} designed a game where agents could donate costly resources only to players that were sufficiently similar to them (given a tag). This was later abstracted into a model in which players can use different payoff matrices (such as those for the Prisoner's Dilemma or the ``Stag Hunt" game) depending on the tag of the opponent~\cite{TraulsenSchuster2003}.  Recognizing another player's identity can in principle be accomplished in two ways: the players can simply record an opponent's tag and select a strategy accordingly~\cite{HammondAxelrod2006b}, or they can try to recognize a strategy by probing the opponent with particular plays. When using tags, it is possible that players cheat by imitating the tag of the opponent~\cite{TraulsenNowak2007} (in that case it is necessary for players to agree on a new tag so that they can continue to reliably recognize each other). 

A tag-based ZD strategy (``ZDt") can cooperate with itself, while playing ZD against non-ZD players. Let us first test that using tags renders ZDt evolutionarily stable against a strategy that ZD loses to, namely All-D. The effective payoff matrix becomes (using the standard payoff values and our example ZD strategy $p_1=0.99,p_4=0.01$)
\be
\bordermatrix{\mbox{} & {\rm ZDt}& {\rm AllD}  \cr
                            {\rm ZDt}  & 3 & 0.75 \cr
                            {\rm AllD}     &2 &  1}
                           \;,\label{ZDt1}
\ee
and we note that now both ZDt and All-D can be an ESS. The game described by the matrix (\ref{ZDt1}) belongs to the class of {\em coordination} games (a typical example is the Stag Hunt game~\cite{Skyrms2004}), which means that the interior fixed point of the dynamics $(\pi_{\rm ZDt},\pi_{\rm AllD})=(0.2,0.8)$ is itself unstable~\cite{Zeeman1980} and the winner of the competition depends on the initial density of strategies. This is a favorable game for ZDt, as it will outcompete AllD as long a its initial density exceeds 20\% of the population. What happens if the opposing strategy acquires the capacity to distinguish self from non-self as well? The optimal strategy in that case would defect against ZDt players, but cooperate with itself. The effective payoff matrix then becomes (``CD" is the conditional defector)
\be
\bordermatrix{\mbox{} & {\rm ZDt}& {\rm CD}  \cr
                            {\rm ZDt}  & 3 & 0.75 \cr
                            {\rm CD}     &2 &  3}
                           \;.\label{ZDt2}
\ee
This game is again in the class of coordination games, but the interior unstable fixed point is now $(\pi_{\rm ZDt},\pi_{\rm CD})=(9/13,4/13)$, which is not at all favorable for ZDt anymore as the strategy now needs to constitute over 69\% of the population in order to drive the conditional defector into extinction. We thus find that tag-based play leads to dominance based on numbers (if players cooperate with their own kind), where a tag-based ZD strategy is only favored if it is the only one that can recognize itself. Indeed, tag-based recognition is used to enhance cooperation among animals via the so-called ``green-beard" effect~\cite{Hamilton1964b,Dawkins1976}, and can give rise to cycles between mutualism and altruism~\cite{Sinervoetal2006}. Recognizing a strategy from behavior rather than from a tag is discussed further below. Note that whether a player's strategy is identified by a tag or learned from interaction, in both cases it is communication that enables cooperation~\cite{Iliopoulosetal2010}.

\subsubsection*{Short-memory players cannot set the rules of the game}

In order to recognize a player's strategy via its actions, it is necessary to be able to send complex sequences of plays, and react conditionally on the opponent's actions. In order to be able to do this, a strategy must be able to use more than just the previous plays (memory-one strategy). This appears to contradict the conclusion reached in~\cite{PressDyson2012} that the shortest-memory player ``sets the rule of the game". This conclusion was reached by correctly noting that in a direct competition of a long-memory player and a short-memory player, the payoff to both players is unchanged if the longer memory player uses a ``marginalized" short-memory strategy. However, as we have seen earlier, in an evolutionary setting it is necessary to not only take cross-strategy competitions into account, but also how the strategies fare when playing against themselves, that is, like-strategies. Then, it is clear that a long-memory strategy will be able to recognize itself (simply by noting that the responses are incompatible with a ``marginal" strategy) and therefore distinguish itself from others. Thus, it appears possible that evolutionarily successful ZD strategies can be designed that use longer memories to distinguish self from non-self. Of course, such strategies will be vulnerable to mutated strategies that look sufficiently like a ZD player so that ZD will avoid them (a form of defensive mimicry~\cite{Malcolm1990}), while being subtly exploited by the mimics instead. 

\section*{Discussion}
ZD strategies are a new class of conditional stochastic strategies for the iterated PD (and likely other games as well) that are able to unilaterally set an opponent's payoff, or else set the ratio of payoffs between the ZD strategist and its opponent. The existence of such strategies is surprising, but they are not evolutionarily (or even mutationally) stable in adapting populations. Evolutionary stability can be determined by using the steady-state payoffs of two players engaged in an unlimited encounter as the one-shot payoff matrix between these strategies. Applying Maynard Smith's standard ESS conditions to that effective payoff matrix shows that ZD strategies can be at most weakly dominant (their payoff against self is equal to what any other strategy receives playing against them), but that it is the opposing strategy that is most often the ESS. (ZDe strategies are not even weakly dominant, except for the limiting case $\chi\to\infty$.) It is even possible that ZD strategies win every single matchup against non-ZD strategies (such as against the strategy ``Pavlov"), yet be evolutionarily unstable and be driven to extinction.

While this argument relies on using the steady-state payoffs, it turns out that an agent-based simulation with finite iterated games reproduces those results almost exactly. Furthermore, ZD strategies are mutationally unstable even when they are the ESS at zero mutation rate, because the proliferation of ZD mutants that are not exactly ZD creates an insurmountable obstacle to the evolutionary establishment of ZD. Rather than taking over, the ZD strategy instead evolves into a harmless cooperating or defecting strategy (depending on the mutation rate, see~\cite{Iliopoulosetal2010}). 

For ZD strategists to stably and reliably outcompete other strategies, they have to have an informational advantage. This ``extra information" can be obtained either by using a tag to recognize each other and conditionally cooperate or play ZD depending on this tag, or else by having a longer-memory strategy that a player can use to probe the opponent's strategy. Tag-based play leads effectively to a game in the ``coordination" class if players cooperate with themselves but not against the opponent, with the winner determined by the initial density.
Of course, such a tag- or an information-based dominance is itself vulnerable to the evolution of interfering mechanisms by the exploited strategies, either by imitating the tag (and thus destroying the information channel) or by evolving longer memories themselves. Indeed, any recognition system that is gene-based is potentially subject to an evolutionary arms race~\cite{DawkinsKrebs1979,Ruxtonetal2004}. In the realm of host-parasite interactions, this evolutionary arms race is known as the Red Queen effect~\cite{vanValen1973} and appears to be ubiquitous throughout the biosphere. 

\section*{Methods}

\subsubsection*{Definition of ZD strategies}
Let $\vec p=(p_1,p_2,p_3,p_4)$ be the probabilities of the ``focal" player to cooperate given the outcomes CC,CD,DC and DD of the previous encounter, and $\vec q=(q_1,q_2,q_3,q_4)$ the probabilities of an opposing strategy (hereafter the ``O"-strategy). Given the payoffs (\ref{pay}) for the four outcomes, the game can be described by a Markov process defined by these eight probabilities~\cite{HauertSchuster1997} (because this is an infinitely repeated game, the probability to engage in the first move--which is unconditional--does not play a role here). 
Each Markov process has a stationary state given by the left eigenvector of the Markov matrix (constructed from the probabilities $\vec p$ and $\vec q$, see~\cite{PressDyson2012}), which in this case describes the equilibrium of the process. The expected payoff is given by the dot product of the stationary state and the payoff vector of the strategy. But while the stationary state is the same for either player, the payoff vector--given by the score received for each of the four possible plays CC, CD, DC, and DD--is different for the two players for the asymmetric plays CD and DC. The mean (that is, expected) payoff for either of these two players calculated in this manner turns out to be a complicated function of 12 parameters: the eight probabilities that define the players' strategies, and the four payoff values in Eq.~(\ref{pay}). The mathematical surprise offered up by Press and Dyson concerns these expected payoffs: it is possible to force the opponent's expected payoff to not depend on $\vec q$ at all, by choosing~\cite{PressDyson2012}
\be
p_2=\frac{p_1(T-P)-(1+p_4)(T-R)}{R-P}\; \;,\; \;p_3=\frac{(1-p_1)(P-S)+p_4(R-S)}{R-P}\;. \label{zdprobs}
\ee
In this case, no matter which strategy O adopts, its payoff is determined entirely by $\vec p$ (the focal player's strategy) and the payoffs. Technically speaking, this is possible because the expected payoff is a linear function of the payoffs for each of the four plays, and as a consequence it is possible for one strategy to enforce the payoff of the opponent by a judiciously chosen set of probabilities that makes the linear combination of determinants vanish (hence the name ``zero determinant" strategies). This enforcement is asymmetric because of the asymmetry in the payoff vectors introduced earlier: while the ZD player can force the opponent's payoff to not depend on their own probabilities, the payoff to the ZD player depends on both the ZD player's as well as the opponent's probabilities. Furthermore, the expected payoff to the ZD opponent (which as mentioned is usually a very complicated function) becomes very simple.

Note that the set of possible ZD strategies is larger than the set studied by Press and Dyson (a two-dimensional subset of the three-dimensional set of ZD strategies, where dimension refers to the number of independently varying probabilities, for example $p_1$ and $p_4$ as in Eq.~(\ref{zdprobs})), but we do not expect that extending the analysis to three dimensions will change the overall results as the third dimension will tend to linearly {\em increase} the payoff of the opposing strategy (rather than keeping it fixed), which benefits the opponent, not the ZD player.

\subsubsection*{Steady-state payoff to the ZD strategist} 
If we set the payoff matrix (\ref{pay}) to the standard values of the Prisoner's Dilemma, i.e., $(R,S,T,P)= (3,0,5,1)$, the payoff  Eq.~(\ref{zd2}) received by the ZD strategist (defined by the pair $p_1$ and $p_4$) playing against an arbitrary stochastic strategy $\vec q$ can be calculated to be

\begin{eqnarray}
&&g(\vec p, \vec q)= \nonumber \\
&&\Bigg\{p_4^2\bigg[q_3(7q_1-1)+3q_2(q_1-5q_3-q_4-1)+2q_4(9q_3-5q_1+2)\bigg] \nonumber \\
&-&p_1^2\bigg[q_2(5q_3+6q_4+2)-2q_3(13q_4+3)+q_1(20q_4-6q_2+q_3+4)\bigg]\nonumber\\
&+&p_4\bigg[q_2(7q_1-14q_3-9q_4-5)+q_3(6q_1+35q_4-5)-2q_4(1+10q_1)-2\bigg]\nonumber\\
&+&p_1\bigg[q_3(8q_2-p_4+20p_4q_2-10)+q_4\Big(p_4(9q_2-44q_3-8)+12q_2-39q_3+20\Big)\nonumber\\
&&\ \ \ +q_1\Big(p_4(30q_4-9q_2-6q_3+2)+2q_3+20q_4-8q_2+4\Big)+7p_4q_2+4\bigg]\nonumber\\
&+&2q_2(1+q_1)+q_3(4-q_1-3q_2)-q_4(20+6q_2-13q_3)-4\Bigg\}/ \nonumber\\
&&\!\!\!\!\!\!\!\!(p_1-p_4-1)\Bigg\{q_3\bigg[q_1+p_4(q_1-3)-4\bigg]+q_4\bigg[p_4(5q_1-6q_3-3)-3q_3+5\bigg]\nonumber\\
&+&q_2\bigg[p_4(q_4+5q_3-6q_1+6)+q_4+3q_3-2q_2-2)\bigg]\nonumber\\
&-&p_1\bigg[q_1(5q_4+q_3-6q_2+4)+q_2(q_4+5q_3+2)-6q_3(1+q_4)\bigg]+4\Bigg\} \label{big}
\end{eqnarray}
Interesting limiting cases are the payoffs against All-D Eq.~(\ref{alld}), as well as the payoff against an unconditional cooperator, given by 
\be
g(\vec p, \vec q=\vec1)=3+\frac43 \left(\frac{1-p_1}{1-p_1+p_4}\right)\;. \label{allc}
\ee
For the strategy Pavlov, using $\vec q_{\rm PAV}=(1,0,0,1)$ in (\ref{big}) yields
\be
g(\vec p, \vec q_{\rm PAV})=\frac{6\left(2(1-p_1)+p_4\right)^2}{\left(9(1-p_1)+2p_4\right)\left(1-p_1+p_4\right))}\;,
\ee
or $g(\vec p, \vec q_{\rm PAV})=27/11$ for the ZD strategy with $p_1=0.99$ and $p_4=0.01$.

\subsubsection*{Agent-based modeling of iterated game play}
In order to study how conditional strategies such as ZD play in finite populations, we perform agent-based simulations in which strategies compete against each other in a well-mixed population of 1,024 agents as in~\cite{Adamietal2012}.  
Every update, an agent plays one move against eight random other agents, and the payoffs accumulate until either of the playing partners is replaced. The fitness of each player is the total payoff accumulated, averaged over all eight players he faced. Because we replace 0.1\% of the population every update using a death-birth Moran process,  on average an agent plays 500 moves against the same opponent (each player survives on average $1/r=1,000$ updates in the population). 
Agents cooperate or defect with probability 0.5 on the first move as this decision is not encoded within the probabilistic set of four (conditional) probabilities. We start populations at fixed mixtures of strategies (for example, ZD versus Pavlov as in Fig.~\ref{fig:sdvszd}), and update the population until one of the strategies has gone to extinction. In such an implementation, there are no mutations and as a consequence strategies do not evolve.
\subsubsection*{Agent-based modeling of strategy evolution}
To simulate the Darwinian evolution of stochastic conditional strategies, we encode the strategy into five loci, encoding the conditional probabilities $\vec q=(q_1,q_2,q_3,q_4)$ as well as the unconditional probability $q_0$. Like in the simulation of iterated game play without mutations described above, agents play eight randomly selected other agents in the (well-mixed) population, and are replaced with a set of genotypes that survived the 1\% removal per update and selected with a probability proportional to their fitness. After the selection process, probabilities are mutated with a probability of 1\% per locus. The mutated probability is drawn from a uniformly distributed random number on the interval $[0,1]$. In order to visualize the course of evolution, we reconstruct the ancestral line of decent (LOD) of a population by retracing the path evolution took backwards, from the strategy with the highest fitness at the end of the simulation back to the ancestral genotype that served as the seed. Because there is no sexual recombination between strategies, each population has a single line of descent after moving past the most recent common ancestor of the population, that is, all individual LODs coalesce to one. The line of descent recapitulates the evolutionary history of that particular run, as it contains the sequence of mutations that gave rise to the successful strategy at the end of the run (see, e.g.,~\cite{Lenskietal2003,Ostmanetal2012}). Because the evolutionary trajectory for any particular locus is usually fairly noisy, we average trajectories over many replicate runs in order to capture the selective pressures affecting each gene. 


\section*{Acknowledgments}
We thank Jacob Clifford for discussions. This work was supported by the National Science Foundation's BEACON Center for the Study of Evolution in Action under  contract No. DBI-0939454. We acknowledge the support of Michigan State University's High Performance Computing Center and the Institute for Cyber Enabled Research (iCER).
\bibliographystyle{naturemag}


\section*{Contributions} CA and AH conceived and designed the study and carried out the research. All authors contributed to writing the manuscript.

\section*{Competing financial interests}
The authors declare no competing financial interests.










\end{document}